\documentstyle[12pt]{article}
\begin{document}
\title{
{\bf New quantum Monte Carlo study of}\\
{\bf quantum critical phenomena with}\\
{\bf Trotter-number-dependent finite-size}\\
{\bf scaling and non-equilibrium relaxation}
}
\author{
{\Large {\sl Yoshihiko Nonomura}}\\
{\large {\it Magnetic materials laboratory, Institute of physical and}}\\
{\large {\it chemical research (RIKEN), Wako, Saitama 351-0108, 
Japan}}\footnote{Present address: 
Computational materials science division, National research 
institute for metals (NRIM), Tsukuba, Ibaraki 305-0047, Japan}
}
\date{{\large PACS numbers: 05.70.Jk, 64.60.Ht, 75.10.Jm}}
\maketitle

\begin{abstract}
We propose a new efficient scheme for the quantum Monte Carlo study 
of quantum critical phenomena in quantum spin systems. Rieger and 
Young's Trotter-number-dependent finite-size scaling in quantum spin 
systems and Ito {\it et al.}'s evaluation of the transition point 
with the non-equilibrium relaxation in classical spin systems are 
combined and generalized. That is, only one Trotter number and one 
inverse temperature proportional to system sizes are taken for each 
system size, and the transition point of the transformed classical 
spin model is estimated as the point at which the order parameter 
shows power-law decay. The present scheme is confirmed by the 
determination of the critical phenomenon of the one-dimensional 
$S=1/2$ asymmetric XY model in the ground state. The estimates of 
the transition point and the critical exponents $\beta$, $\gamma$ and 
$\nu$ are in good agreement with the exact solutions. The dynamical 
critical exponent is also estimated as ${\mit \Delta}=2.14\pm 0.06$, 
which is consistent with that of the two-dimensional Ising model. 
\end{abstract}

\section{Introduction}
Quantum critical phenomena are one of the most interesting topics 
in low-dimensional quantum spin systems, and various numerical methods 
have been proposed. In one-dimensional systems, the exact-diagonalization 
method gives the most reliable result when the chain is single and the 
size of spins is not so large. In multi-leg ladder systems or large-spin 
systems, reachable system sizes with this method are not satisfactory and 
other approximate methods should be used. Historically, the quantum Monte 
Carlo (QMC) method\cite{qmc,qmcrev} and related methods (the quantum 
transfer-matrix method\cite{qmcrev,qtr1,qtr2}, the Monte Carlo power 
method\cite{qmcrev,mpow} {\it etc}.) have been used, and nowadays the 
density-matrix renormalization-group (DMRG) method\cite{dmrg1,dmrg2,pwfrg} 
is considered to be most powerful because quite large (effectively 
infinite) systems can be calculated with small errors and relatively 
small required CPU demand. 
The DMRG method is applicable even to two-dimensional systems with 
large enough energy gap, but the limitation of computer memories makes 
the estimation of critical phenomena difficult at present (any systems 
are gapless at the critical point in the thermodynamic limit). 
Then, the QMC method is still most useful to study quantum critical 
phenomena in non-frustrated two-dimensional quantum spin systems. 

In QMC simulations of ground states in quantum spin systems, the 
most serious problem is large sampling errors. The QMC method 
is essentially applicable only to finite-temperature systems, 
and ``ground states" mean nothing but ``calculations at low enough 
temperatures", which might include ambiguity. Moreover, simulations 
at low temperatures result in large correlation time, which is the 
origin of large sampling errors. Furthermore, in the conventional 
world-line QMC method, physical quantities are evaluated after the 
extrapolation of the data for various Trotter numbers, which makes 
statistical errors larger. 
These shortcomings of the world-line QMC algorithm have been overcome 
with the stochastic series expansion method\cite{sse1,sse2,sse3}, 
the quantum cluster algorithm\cite{clqmc,contim} or Prokof'ev's 
algorithm\cite{proko,promod} proposed recently. The Trotter 
extrapolation is not required in these methods, and fast relaxation 
of them makes simulations at very low temperatures possible. 
As long as finite systems are considered, energy gaps are finite even at the 
transition point, and simulations at the temperatures lower enough than 
energy gaps give ``ground-state" properties with convergent correlation time. 
Although such a brute-force strategy is required for the evaluation of actual 
physical quantities at the ground state, more efficient scheme is expected 
to study quantum critical phenomena, in which universal behavior exists. 

In the present paper, we propose a new efficient scheme to study 
quantum critical phenomena in quantum spin systems with the QMC method. 
This scheme can be regarded as a combination and a generalization of 
two independent techniques of Monte Carlo calculations originally 
proposed in Ising spin systems. That is, Rieger and Young's 
Trotter-number-dependent finite-size scaling\cite{triqsg2d} introduced 
in the QMC study of the quantum critical phenomenon of the Ising spin 
glass model in a transverse field is coupled with Ito {\it et.\ al.}'s 
determination of the transition point\cite{noneqdec} based on the 
non-equilibrium relaxation of the order parameter. The latter method 
was introduced in the classical Ising spin glass model, and the former 
method is necessary for the generalization to quantum spin systems. 
By use of the present scheme, only one Trotter number has to be 
considered in each system size for the determination of the universality 
class. In short, the combination of these two methods is essential 
for efficient numerical calculations in quantum spin systems. 

In order to verify the applicability of the present scheme, we estimate 
the transition point and critical exponents of the one-dimensional $S=1/2$ 
asymmetric XY model. This model is mapped on a two-dimensional Ising spin 
model with conservation laws (or infinite multi-body interactions in the 
Ising-spin representation) by the Suzuki-Trotter decomposition\cite{qmc}, 
and the universality class of this model\cite{xycri1,xycri2,xycri3} is 
different from that of the two-dimensional Ising model. In section 2, 
the present scheme is explained together with a brief review of the 
world-line QMC method. In section 3, the quantum critical phenomenon 
of the one-dimensional $S=1/2$ asymmetric XY model is analyzed. The 
transition point and the dynamical critical exponent are estimated with 
very large clusters (linear sizes are $L=1024$, $2048$ and $4096$) and 
various critical exponents are estimated with the clusters up to $L=128$. 
The above descriptions are summarized in section 4. Part of the present 
results have already been briefly reported elsewhere\cite{newqmclet}. 

\section{Numerical methods}
In the world-line quantum Monte Carlo method\cite{qmc,qmcrev}, 
the partition function of quantum spin systems is transformed 
to a classical one by the Suzuki-Trotter formula, 
\begin{equation}
  Z\equiv{\rm Tr}\hspace{0.5mm}{\rm e}^{-\beta{\cal H}}
        =\lim_{M\to\infty}{\rm Tr}
         \left[\exp\left(-\frac{\beta}{M}{\cal H}_{1}\right)
               \exp\left(-\frac{\beta}{M}{\cal H}_{2}\right)\right]^{M},
\end{equation}
where $\beta$ stands for the inverse temperature ($\beta\equiv 1/T$), 
and the partial Hamiltonians ${\cal H}_{1}$ and ${\cal H}_{2}$ in 
one-dimensional systems are shown in figure \ref{decfig}, which is known 
to be the checkerboard decomposition. Then, each 4-spin unit cell is 
independent, and Monte Carlo simulations are possible. Conventional 
local flips and the $x$-direction global flip are introduced. When 
the models have conservation laws, the Trotter-direction global flip 
should be considered for the evaluation of the thermal averages in the 
whole phase space. As is well known, the tangent-direction global flip 
can be safely omitted, and this algorithm is vectorized\cite{qmcvec}. 

In the conventional world-line QMC method, quantities 
in the ground state are evaluated as follows: 
\begin{enumerate}
\item The temperature is fixed at a ``low enough" one, $T_{0}(L)$, which 
is much smaller than the energy gap of finite clusters, $\Delta E(L)$, 
in the subspace in which the ground state is included. 
The Trotter-direction global flip is not introduced in this case. 
\item QMC simulations are performed for some finite Trotter numbers 
$M$, and quantities in the $M\to\infty$ limit are estimated with the 
following scaling form, $Q(M,L)=Q(\infty,L)+{\rm const.}/M^{2}+\cdots$. 
\item Quantities in the thermodynamic limit are evaluated from the 
finite-size scaling of $\{Q(\infty,L)\}$ for various system sizes. 
\end{enumerate}
In the above analysis, the $T\to 0$ limit is skipped by use of the finite 
energy gap in finite systems, but double extrapolations with respect to 
$M$ and $L$ are still necessary. Calculations for some Trotter numbers 
are required for each system size, and simulations at a ``low enough" 
temperature consume much CPU demand because of long correlation time. 

Although the above analysis is required for the evaluation of a certain 
physical quantity in the ground state, such a complicated procedure is 
not necessary for the investigation of quantum critical phenomena. 
The universality class of the original $D$-dimensional quantum 
system in the ground state is expected to be equivalent to 
that of the transformed $(D+1)$-dimensional classical system, 
and Rieger and Young\cite{triqsg2d} pointed out that the finite-size 
scaling holds when the ratios $M/L^{z}$ and $\beta/M$ are fixed. 
The exponent $z$ stands for the ratio of the correlation length 
in the real direction ($\xi$) and that in the Trotter direction 
($\xi_{\tau}$), namely $\xi^{z}\propto \xi_{\tau}$ 
in the vicinity of the transition point. This exponent 
is given by $z=1$ in non-random systems (the Lorentz invariance). 
In this scheme, the temperature is scaled in accordance with the 
system size, and the ambiguity of the simulated temperature 
is automatically removed. Moreover, this temperature can be 
higher than the standard ``low enough'' temperature. 

The transition point has been estimated from physical quantities in 
equilibrium until present. When it is determined by the finite-size 
scaling analysis, the critical exponent should also be evaluated 
simultaneously, and the accuracy of estimates becomes poor. 
When it is estimated from scale-invariant quantities, the 
above disadvantage is excluded. However, statistical errors of 
such complicated quantities are larger than those of ordinary 
quantities, and the final estimates are not expected to be 
so accurate again. Then, in the present paper, we utilize a 
new method for the determination of the transition point recently 
proposed by Ito {\it et.\ al.}\cite{noneqdec}. They showed that 
the transition point can be estimated quite accurately with 
the non-equilibrium relaxation of the order parameter. 
That is, simulations are started from a perfectly-ordered state, 
and the order parameter is measured in each Monte Carlo step (MCS). 
The order parameter decays exponentially in the disordered phase, 
remains non-vanishing in the ordered phase, and power-law decay 
is observed only at the transition point (figure \ref{decschfig}). 
This calculation uses only several hundreds or a few thousands of 
MCS in each sample, and very large systems can be easily treated 
without equilibration and Monte Carlo sampling. 

Since the Trotter extrapolation is justified only in equilibrium 
quantities, this method is not consistent with the conventional 
world-line QMC method. Thus, the combination with Rieger and Young's 
scheme is indispensable for applications to quantum spin systems. 
Since the transition point is not a universal value, the estimate 
obtained from transformed classical spin models with fixed $\beta/M$ 
is generally different from the transition point of the original 
quantum spin model. However, the transition point of the transformed 
model should be used in the finite-size-scaling analysis of a series of 
systems with fixed $\beta/M$, and the ``true" transition point of 
the original model is estimated from some simulations with 
different values of $\beta/M$. Details of this analysis 
will be explained elsewhere\cite{afdim2d}. 

The present scheme is so general that it can be coupled with any kinds 
of Monte Carlo dynamics in principle. Although values of the dynamical 
critical exponent depend on the dynamics (details will be explained in 
the next section), the power-law-decay behavior at the transition point 
is common as long as second-order phase transitions are considered. 
The essential point of the usage of Rieger and Young's scheme in the 
present framework is the idea that simulated temperatures are scaled 
in accordance with variance of system sizes. Thus, Trotter-number-free 
methods such as the continuous-time loop algorithm\cite{contim} 
can be utilized. On that occasion, $\beta/L^{z}$ should be fixed 
instead of $M/L^{z}$ and $\beta/M$. 
On the other hand, the accuracy of the estimate of the transition 
point depends on the dynamics. If a very fast dynamics such as the loop 
algorithm is utilized, the decay process is completed within first several 
MCS and precise evaluation of the transition point would be difficult 
because of large fluctuations of the order parameter and the discontinuity 
of MCS. Of course, this statement is also model dependent: even the loop 
algorithm might be useful in systems with principal slow dynamics such as 
quantum spin glass models. We can at least tell that the conventional 
``slow" dynamics is suitable for regular systems as will be shown 
in the next section. 

\section{Numerical calculations of critical phenomenon}
In order to verify the applicability of the new scheme explained 
in the previous section, we analyze the quantum critical phenomenon 
of the one-dimensional $S=1/2$ asymmetric XY model described by 
the following Hamiltonian, 
\begin{equation}
  \label{xyham}
  {\cal H}=-\sum_{i=1}^{L}S_{i}^{x}S_{i+1}^{x}
           -g\sum_{i=1}^{L}S_{i}^{y}S_{i+1}^{y}
           -H^{x}\sum_{i=1}^{L}S_{i}^{x}\ ,
\end{equation}
with the periodic boundary condition ($S_{L+1}^{x,y}=S_{1}^{x,y}$). 
The in-plane field $H^{x}$ is introduced only for the definition of the 
zero-field susceptibility, and it is not applied in actual simulations. 
We only consider even-spin systems which have a singlet ground state. 
This is the simplest quantum spin model with conservation laws, and 
this model has already been solved exactly\cite{xycri1,xycri2,xycri3}. 
There are no matrix elements between the two subspaces with 
$\sum_{i}S_{i}^{x}={\it odd}$ and $\sum_{i}S_{i}^{x}={\it even}$. 
The apparent transition point $g_{{\rm c}}=1$ is common in the 
original model (\ref{xyham}) and the transformed classical spin models. 
Simulations are based on the up-down basis along the $x$-direction, and 
the Trotter number and simulated temperature are fixed as $2M/L=1$ and 
$\beta/M=1$, following Rieger and Young's parameterization\cite{triqsg2d}. 
Under these conditions, the original $L$-spin system is transformed to 
a classical $L\times L$-spin system with all the interactions unity at 
$g=1$, and good scaling behavior is expected. The phase space is limited 
to the one with $\sum_{i}S_{i}^{x}={\it even}$ in which the ground 
state is included. Since finite (not ``low enough'') temperatures 
are introduced in the present scheme, the above restriction of the 
phase space cannot be justified a priori. Complementary simulations 
without this restriction are discussed in Appendix. 

\subsection{Transition point $g_{{\rm c}}$} 
First, the transition point is estimated with the non-equilibrium relaxation 
of the order parameter. When the $x$-component of the magnetization, 
\begin{equation}
  m^{x}(L)\equiv\frac{1}{L}\sum_{i=1}^{L}S_{i}^{x}\ ,
\end{equation}
is regarded as the order parameter, the regions $g<1$ and $g>1$ correspond 
to the ordered and disordered phases, respectively. Simulations are started 
from the all-up state, and the initial $3\times 10^{3}$ Monte Carlo 
steps (MCS) are measured for the $L=1024,2048$ and $4096$ systems. 
In simulations for these system sizes, data with $20$, $5$ and $1$
independent sequences of random numbers are averaged, respectively. 
The order parameter of the $L=4096$ system is plotted versus MCS in a 
logarithmic scale in figure \ref{decsimfig}(a), and this behavior is the 
same as displayed in figure \ref{decschfig}. That is, the data at $g=1.000$ 
are almost on a straight line, and the data at $g=0.996$ and $g=1.004$ 
are clearly upward- and downward-bending, respectively. Therefore, the 
transition point is estimated as $g_{{\rm c}}=1.000\pm 0.002$ within the 
present simulations, as expected. The data for the $L=1024$ and $2048$ 
systems at $g=1.000$ are given in figure \ref{decsimfig}(b) together 
with those for the $L=4096$ system, and this figure means that there 
exists no explicit systematic size dependence in such large systems. 
Since values of $L\times 2M$ spins are averaged in these simulations, 
the data show relatively smooth behavior in spite of small number of 
averaged random-number sequences. 

Next, we perform similar simulations in smaller systems. In equilibrium 
calculations, the ``approximate" transition point $g_{{\rm ap}}(L)$ 
such as the maximum of the specific heat shows the following systematic 
deviation from the transition point $g_{{\rm c}}$ in finite systems, 
\begin{equation}
  \label{pointsc}
  g_{{\rm ap}}(L)=g_{{\rm c}}+{\rm const.}\times L^{1/\nu}+\cdots\ ,
\end{equation}
and the critical exponent of the correlation length, $\nu$, is estimated. 
The relaxation of the order parameter for $L=32$ and $64$ (the numbers 
of averaged samples are $16384$ and $4096$, respectively) is displayed 
in figures \ref{decsmlfig}(a) and \ref{decsmlfig}(b), respectively. 
These figures show that the order parameter decays exponentially 
after a certain MCS even for $g<1$. This property is quite natural 
because the order parameter vanishes regardless of the value of $g$ 
in finite systems in equilibrium. These figures also reveal that 
the crossover MCS from ``quasi-infinite" properties to finite-size 
ones depends on system sizes, namely $\sim 40$ MCS for $L=32$, and 
$\sim 180$ MCS for $L=64$. These results suggest that the above 
behavior of $L>1000$ systems is nothing but transient, but that 
the present $3\times 10^{3}$ MCS are small enough to be regarded 
as ``quasi-infinite" time scale. Moreover, the transition point 
estimated from these figures is also $g_{{\rm c}}\simeq 1.00$, and 
the size dependence of the estimates seems quite small. This behavior 
is not unphysical because the ``quasi-infinite" property means that 
finite-size effects are not apparent in this time scale. Thus, the 
scaling form (\ref{pointsc}) is not useful for the estimation 
of the critical exponent $\nu$ within the present scheme. 

As explained in the previous section, the ``transition point" estimated 
with the present scheme generally depends on the value of $\beta/M$. 
On the other hand, it is independent of the value of $M/L$, because 
``quasi-infinite" properties are observed in the present scheme and the 
difference of the shape of transformed clusters from squares is expected 
to result only in the difference the crossover MCS as shown above. 
Then, in order to confirm this argument, we simulate systems with 
(a) $4M/L=1$ ($L=4096$, $M=1024$, $2$ samples) and (b) $M/L=1$ 
($L=2048$, $M=2048$, $2$ samples). These results are plotted versus 
MCS in a logarithmic scale in figures \ref{diffig}(a) and \ref{diffig}(b), 
respectively. These figures also give $g_{{\rm c}}=1.000\pm 0.002$, 
as expected. 

\subsection{Dynamical critical exponent ${\mit \Delta}$}
According to the dynamical finite-size scaling theory\cite{dynfss1,dynfss2}, 
the non-equilibrium relaxation of the order parameter at the transition 
point is scaled\cite{neqorg1,neqorg2} as 
\begin{equation}
  \label{dynexpsc}
  m^{x}(t)\sim t^{-\lambda}\ ,\ \ \lambda=\frac{\beta}{{\mit \Delta}\nu}\ ,
\end{equation}
where $t$ stands for MCS, the critical exponents $\beta$ and $\nu$ will 
be defined in the next subsection ((\ref{betadef}) and (\ref{nudef})), 
and the dynamical critical exponent ${\mit \Delta}$ is defined as 
\begin{eqnarray}
  \label{dyndef1}
  &&\langle S_{i}^{x}(0)S_{i}^{x}(t)\rangle-\langle S_{i}^{x}(0)\rangle^{2}
        \sim\exp(-t/\tau)\ ,\\
  &&\tau\sim|g-g_{{\rm c}}|^{-{\mit \Delta}}
            \ \ \ {\rm for}\ \ \ g\to g_{{\rm c}}\ .
  \label{dyndef2}
\end{eqnarray}
Although the transition point can be estimated accurately enough with 
the data displayed in figure \ref{decsimfig}(a), these data are too 
divergent to evaluate the exponent $\lambda$ with the local-exponent 
method originally proposed by Stauffer\cite{neqorg1}. Then, we naively 
fit the data shown in figure \ref{decsimfig}(a) with the scaling form 
(\ref{dynexpsc}) for various time scales ($t=t_{{\rm ini}}\sim t_{{\rm fin}}$, 
$\Delta t\equiv t_{{\rm fin}}-t_{{\rm ini}}=50$, $100$ and $200$). 
Estimates of the exponent $\lambda$ are plotted versus 
$1/t_{{\rm mid}}$ in figure \ref{dynexpfig}, 
with $t_{{\rm mid}}\equiv (t_{{\rm ini}}+t_{{\rm fin}})/2$. 
In addition to the data given in figure \ref{decsimfig}(a), results of 
two more independent simulations are averaged in order to obtain 
reliable estimates. In each $\Delta t$, $\lambda$ almost remains 
constant in the early stage, and it begins to fluctuate as $t_{{\rm mid}}$ 
becomes larger. As for the final estimate $\lambda_{{\rm est}}$, 
we require that the central value should be close to the early-stage 
one, and that initial up-down fluctuations should be included 
within the error bar. These conditions are satisfied when we take 
$\lambda_{{\rm est}}=0.1167\pm 0.0033$ (see figure \ref{dynexpfig}). 
Combining this exponent with the exact solution, 
$\beta/\nu=0.25$\cite{xycri1,xycri2,xycri3}, the dynamical 
critical exponent of the present model is estimated as 
\begin{equation}
{\mit \Delta}=2.14\pm 0.06\ ,
\end{equation}
which is consistent with that of the two-dimensional Ising model, 
${\mit \Delta}=2.165\pm 0.010$\cite{neqorg3}. This result might suggest 
that the dynamical critical exponent is common in the free-fermion 
models, even though the universality classes are different. 
Actually, the critical exponent $\nu$ defined by 
\begin{eqnarray}
  \label{nudef1}
  &&\langle S_{i}^{x}S_{j}^{x}\rangle-\langle S_{i}^{x}\rangle^{2}
       \sim\exp(-|r_{i}-r_{j}|/\xi)\ ,\\
  &&\xi\sim|g-g_{{\rm c}}|^{-\nu}\ \ \ {\rm for}\ \ \ g\to g_{{\rm c}}\ ,
  \label{nudef2}
\end{eqnarray}
is unity in both the present model and the two-dimensional Ising model. 
When the definition of ${\mit \Delta}$, (\ref{dyndef1}) and (\ref{dyndef2}), 
is compared with that of $\nu$, (\ref{nudef1}) and (\ref{nudef2}), the 
exponent ${\mit \Delta}$ can be regarded as the MCS version of $\nu$. 
Since these two models have the common $\nu$, it is not strange that they 
might also have the common ${\mit \Delta}$. Of course, the statistical 
error of the present estimate is not small enough, and further studies 
in this direction (averaging much sequences of random numbers or 
simulating larger systems) are required. 

\subsection{Critical exponents $\beta/\nu$ and $\gamma/\nu$}
Next, we estimate the critical exponents which can be obtained 
from equilibrium simulations at $g=g_{{\rm c}}=1$. We evaluate 
the squared magnetization defined by 
\begin{equation}
  m^{2}(g,L)\equiv\left\langle\left(m^{x}(L)\right)^{2}\right\rangle
                 =\frac{1}{2ML^{2}}\sum_{n=1}^{2M}
                  \left\langle\left(\sum_{i=1}^{L}S_{i,n}^{x}\right)^{2}
                              \right\rangle,
\end{equation}
and the zero-field susceptibility for the in-plane field defined by 
\begin{equation}
  \chi(g,L)\equiv\left.\frac{\partial}{\partial H^{x}}
                       \langle m^{x}(L)\rangle\right|_{H^{x}=0}
                =\frac{\beta}{4M^{2}L}
                 \left\langle\left(\sum_{i=1}^{L}\sum_{n=1}^{2M}
                                   S_{i,n}^{x}\right)^{2}\right\rangle.
\end{equation}
Here $n$ stands for the suffix in the Trotter direction, and these 
quantities show the following scaling behavior at the transition point, 
\begin{equation}
  \label{scq1}
  m^{2}(g_{{\rm c}},L)\sim L^{-2\beta/\nu}\ ,\ \ 
  \chi (g_{{\rm c}},L)\sim L^{\gamma/\nu}\ ,
\end{equation}
where the critical exponents $\beta$, $\gamma$ and $\nu$ are defined as 
\begin{eqnarray}
  \label{betadef}
  m(g,\infty)   &\sim&(g_{{\rm c}}-g)^{\beta}
                      \ \ \ \ \ {\rm for}\ \ \ g\to g_{{\rm c}}\ ,\\
  \chi(g,\infty)&\sim&|g-g_{{\rm c}}|^{-\gamma}
                      \ \ \ {\rm for}\ \ \ g\to g_{{\rm c}}\ ,\\
  \xi(g,\infty) &\sim&|g-g_{{\rm c}}|^{-\nu}
                      \ \ \ {\rm for}\ \ \ g\to g_{{\rm c}}\ .
  \label{nudef}
\end{eqnarray}
Simulations are performed for the $L=16$, $24$, $32$, $48$, $64$, 
$80$, $96$, $112$ and $128$ systems, and the MCS for each system 
size are listed in table \ref{mcstab}. Results of these calculations 
are given in table \ref{res1tab} and figures \ref{datafig}(a) and 
\ref{datafig}(b). The critical exponents are estimated by the 
least-squares fitting of the data for $64\leq L\leq 128$ 
with the scaling forms (\ref{scq1}) as 
\begin{equation}
  \label{expest}
  \beta/\nu=0.241\pm 0.003\ ,\ \ \gamma/\nu=1.499\pm 0.005\ .
\end{equation}
On the other hand, when we utilize the following scaling forms 
in which the next-order correction terms are considered, 
\begin{eqnarray}
  \label{scmag2}
  m^{2}(g_{{\rm c}},L)
  &=&a_{1}L^{-2\beta/\nu}+a_{2}L^{-2\beta/\nu-1}+\cdots\ ,\\
  \chi(g_{{\rm c}},L)
  &=&b_{1}L^{\gamma/\nu} +b_{2}L^{\gamma/\nu-1} +\cdots\ ,
  \label{scsus2}
\end{eqnarray}
we have 
\begin{equation}
  \label{weakexp}
  \beta/\nu=0.245\pm 0.006\ ,\ \ \gamma/\nu=1.49\pm 0.02\ ,
\end{equation}
from the least-squares fitting of the data for $24\leq L\leq 128$. 
The estimates (\ref{expest}) and (\ref{weakexp}) are both 
consistent with the exact values, $\beta/\nu=0.25$ and 
$\gamma/\nu=1.5$\cite{xycri1,xycri2,xycri3}. 
\begin{table}[t]
\caption{
Monte Carlo steps for equilibration (E-MCS) and for measurement 
(M-MCS) are listed for various system sizes. 
}
\label{mcstab}
\begin{center}
\item[]\begin{tabular}{rcc}
\noalign{\hrule height 1pt}
  $L$ & E-MCS & M-MCS \\ 
\hline
 $16$ & $0.2\times 10^{6}$ & $0.2\times 10^{7} $ \\
 $24$ & $0.4\times 10^{6}$ & $0.4\times 10^{7} $ \\
 $32$ & $0.8\times 10^{6}$ & $0.8\times 10^{7} $ \\
 $48$ & $1.2\times 10^{6}$ & $1.2\times 10^{7} $ \\
 $64$ & $0.5\times 10^{6}$ & $0.5\times 10^{7} $ \\
 $80$ & $0.8\times 10^{6}$ & $0.8\times 10^{7} $ \\
 $96$ & $1.0\times 10^{6}$ & $1.0\times 10^{7} $ \\
$112$ & $1.2\times 10^{6}$ & $1.2\times 10^{7} $ \\
$128$ & $2.0\times 10^{6}$ & $2.0\times 10^{7} $ \\
\noalign{\hrule height 1pt}
\end{tabular}
\end{center}
\end{table}
\begin{table}[t]
\caption{
Ground-state averages of the squared magnetization and the 
susceptibility at the transition point for various system sizes. 
}
\label{res1tab}
\begin{center}
\item[]\begin{tabular}{rcc}
\noalign{\hrule height 1pt}
  $L$ & $m^{2}$ & $\chi$ \\ 
\hline
 $16$ & $0.11094\pm 0.00031$ 
      & $1.2242   \times 10^{1}\pm 3.0\times 10^{-2}$ \\
 $24$ & $0.09379\pm 0.00035$ 
      & $2.2785   \times 10^{1}\pm 6.5\times 10^{-2}$ \\
 $32$ & $0.08279\pm 0.00037$ 
      & $3.531\ \ \times 10^{1}\pm 1.2\times 10^{-1}$ \\
 $48$ & $0.06868\pm 0.00023$ 
      & $6.589\ \ \times 10^{1}\pm 3.0\times 10^{-1}$ \\
 $64$ & $0.06052\pm 0.00029$ 
      & $1.0299   \times 10^{2}\pm 6.3\times 10^{-1}$ \\
 $80$ & $0.05437\pm 0.00021$ 
      & $1.4385   \times 10^{2}\pm 7.2\times 10^{-1}$ \\
 $96$ & $0.04982\pm 0.00026$ 
      & $1.890\ \ \times 10^{2}\pm 1.3\times 10^{0}\ \ $ \\
$112$ & $0.04624\pm 0.00026$ 
      & $2.381\ \ \times 10^{2}\pm 1.7\times 10^{0}\ \ $ \\
$128$ & $0.04336\pm 0.00022$ 
      & $2.912\ \ \times 10^{2}\pm 1.9\times 10^{0}\ \ $ \\
\noalign{\hrule height 1pt}
\end{tabular}
\end{center}
\end{table}

\subsection{Critical exponent $\nu$}
The critical exponents $\beta/\nu$ and $\gamma/\nu$ are 
related\cite{dynrel} to one another through the exponent $z$ as 
\begin{equation}
  2\beta/\nu+\gamma/\nu=d+z\ ,
\end{equation}
where $d$ denotes the spatial dimension. 
Then, the critical exponent $\nu$ should be estimated independently 
for the complete determination of the universality class. 
This calculation requires several simulations out of the transition point, 
and they consume much more CPU time than the simulations in the 
previous subsection. Since the direct evaluation of the correlation 
length is not so accurate, we instead estimate the critical exponent 
$\nu$ on the basis of the following scaling functions, 
\begin{eqnarray}
  m^{2}(g,L)&\sim&L^{-2\beta/\nu}f(L/\xi(g))
             \sim L^{-2\beta/\nu}F(L^{1/\nu}|g-g_{{\rm c}}|)\ ,\\
  \chi(g,L) &\sim&L^{\gamma/\nu} g(L/\xi(g))\ \ \ 
             \sim L^{\gamma/\nu} G(L^{1/\nu}|g-g_{{\rm c}}|)\ .
\end{eqnarray}
These formulas mean that the derivatives with respect to $g$ behave as 
\begin{equation}
  \label{derqexp}
  \left.\frac{{\rm d}}{{\rm d}g}m^{2}(g,L)\right|_{g=g_{{\rm c}}}
  \sim L^{(1-2\beta)/\nu}\ ,\ \ 
  \left.\frac{{\rm d}}{{\rm d}g}\chi(g,L)\right|_{g=g_{{\rm c}}}
  \sim L^{(1+\gamma)/\nu}.
\end{equation}
Practically, such derivatives are replaced by the evaluation of the 
slopes of the physical quantities just below the transition point. 
For example, the susceptibility of the $L=128$ system is 
displayed in figure \ref{slopefig}. The exponents 
appeared in (\ref{derqexp}) are estimated as 
\begin{equation}
  \label{slopeexp}
  (1-2\beta)/\nu=0.525\pm 0.007\ ,\ \ (1+\gamma)/\nu=2.506\pm 0.006\ ,
\end{equation}
by the least-squares fitting of the data for $80\leq L\leq 128$ 
(see table \ref{res2tab} and figures \ref{slopefitfig}(a) 
and \ref{slopefitfig}(b)) with (\ref{derqexp}). From the 
exponents (\ref{expest}) and (\ref{slopeexp}), we have 
\begin{equation}
  1/\nu=1.007\pm 0.008\ \ {\rm or}\ \ \nu=0.993\pm 0.008\ ,
\end{equation}
assuming the statistical independence of the exponents 
(\ref{expest}) and (\ref{slopeexp}). This estimate is 
consistent with the exact value, $\nu=1$\cite{xycri3}. 
Note that the complete coincidence of the estimates of $\nu$ 
obtained from two different quantities would be by chance. 
\begin{table}[t]
\caption{
Slopes of the squared magnetization and the susceptibility 
just below the transition point for various system sizes. 
}
\label{res2tab}
\begin{center}
\item[]\begin{tabular}{rcc}
\noalign{\hrule height 1pt}
  $L$ & $m^{2}$ & $\chi$ \\ 
\hline
 $64$ & $0.5942\pm 0.0039$ & $1301\pm\ 8$ \\
 $80$ & $0.6804\pm 0.0062$ & $2320\pm 21$ \\
 $96$ & $0.7590\pm 0.0086$ & $3720\pm 42$ \\
$112$ & $0.8191\pm 0.0086$ & $5457\pm 57$ \\
$128$ & $0.8714\pm 0.0074$ & $7567\pm 64$ \\
\noalign{\hrule height 1pt}
\end{tabular}
\end{center}
\end{table}

\section{Summary and future problems}
In the present paper, we have proposed a new efficient finite-size 
scaling scheme to study quantum critical phenomena in quantum spin 
systems on the basis of the quantum Monte Carlo method. Instead of 
the conventional Trotter extrapolation at low enough temperatures, 
we concentrate on transformed Ising spin systems with $2M/L=1$ and 
$\beta/M=1$. The transition point of the transformed model is 
estimated with the non-equilibrium relaxation of the order parameter, 
and various critical exponents are evaluated from the standard 
finite-size scaling of transformed classical spin systems. 
That is, the critical exponents divided by the critical exponent 
of the correlation length, $\nu$, are estimated from the size 
dependence of physical quantities at the transition point. 
The exponent $\nu$ itself is evaluated from the size dependence 
of the slopes of physical quantities just below the transition 
point. Then, the universality class of the models are completely 
determined within the present framework. In order to verify the 
applicability of the present scheme, we have estimated the 
transition point and the critical exponents $\beta$, $\gamma$ 
and $\nu$ of the one-dimensional $S=1/2$ asymmetric XY model, and 
all the critical exponents have been evaluated with high accuracy 
in comparison with the exact solutions. The dynamical critical 
exponent has also been estimated as ${\mit \Delta}=2.14\pm 0.06$, 
which is consistent with that of the two-dimensional Ising model. 
The present estimation of the transition point is more efficient 
than any other methods based on equilibrium quantities. The present 
finite-size-scaling scheme is so general that it can be combined 
with improved algorithms such as the quantum cluster algorithm. 

The present simulations have been performed in the restricted phase 
space with $\sum_{i}S_{i}^{x}={\it even}$ in which the ground state 
is included. Complementary simulations without this restriction 
(simulations including the Trotter-direction global flip) have also 
been performed, and we have found that theses simulations require 
much CPU demand (about 1.3 times larger in the present model) and 
show larger finite-size corrections than those with the restriction 
of the phase space. These results mean that simulations including 
the Trotter-direction global flip are useless. We emphasize that the 
conventional ``slow" single-spin-flip-type dynamics is practically useful 
for the precise estimation of the transition point in the present scheme. 
On the other hand, the introduction of the quantum cluster algorithm 
in simulations of equilibrium quantities is an important future 
task for us. Applications to quantum critical phenomena of the 
two-dimensional $S=1/2$ dimerized Heisenberg models\cite{afdim2d} and 
the two-dimensional $S=1/2$ random Heisenberg models\cite{afran2d} 
will be reported in the near future. 

\section*{Acknowledgements}
The present author is supported by the Special Researcher's Basic Science 
Program from RIKEN. He would like to thank Professor S.~Miyashita for 
valuable suggestions and useful comments. He used the vectorized 
random-number generation program RNDTIK developed by Professor N.~Ito 
and Professor Y.~Kanada. Most numerical calculations were performed on 
FACOM VPP500 at the Institute for Solid State Physics, University of Tokyo. 

\section*{Appendix}
Here we show the results of simulations including the Trotter-direction 
global flip. Since details of the analysis have already been explained 
in section 2 and 3, only the results for the transition point and the 
critical exponents $\beta/\nu$ and $\gamma/\nu$ are briefly summarized. 

The transition point is estimated with the $L=4096$ system for one 
sequence of random numbers. The decay of the order parameter from 
the all-up state is measured during $3\times 10^{3}$ MCS as shown in 
figure \ref{decgffig}, and we have $g_{{\rm c}}=1.000\pm 0.002$. 
Although this estimate is the same as obtained in section 3, 
fluctuations of the order parameter become larger than those 
measured in section 3. This result means that the present 
scheme also works well when the Trotter-direction global 
flip is included, but the expansion of the phase space 
results in the increase of fluctuations, which is not 
desirable for accurate estimation of critical phenomena. 

The squared magnetization and the zero-field susceptibility are 
simulated at the transition point $g=g_{{\rm c}}=1$ for the same 
system sizes and the same MCS as listed in table \ref{mcstab}. 
The size dependence of these quantities are displayed in 
figures \ref{scgffig}(a) and \ref{scgffig}(b), and the critical 
exponents are estimated with the scaling forms (\ref{scq1}) as 
\begin{equation}
  \label{expestgf}
  \beta/\nu=0.219\pm 0.007\ ,\ \ \gamma/\nu=1.546\pm 0.009\ ,
\end{equation}
from the $64\leq L\leq 128$ clusters, and with 
(\ref{scmag2}) and (\ref{scsus2}), we have 
\begin{equation}
  \label{expestgf2}
  \beta/\nu=0.22\pm 0.03\ ,\ \ \gamma/\nu=1.56\pm 0.05\ ,
\end{equation}
from the $32\leq L\leq 128$ clusters. These estimates are 
not inconsistent with the exact solutions, $\beta/\nu=0.25$ and 
$\gamma/\nu=1.5$, but the accuracy is not so good as the results 
obtained in section 3 due to larger finite-size corrections. 

The above two numerical calculations with respect to non-equilibrium 
relaxation and equilibrium quantities are based on the same algorithm, 
and they consume larger CPU time than those in section 3 because of 
the extra Trotter-direction global flip. They spend about $1.3$ 
times larger CPU demand in the present model. All the results 
suggest that simulations including the Trotter-direction global 
flip are inferior to those without that kind of global flip.

\section*{Figure captions}
\noindent
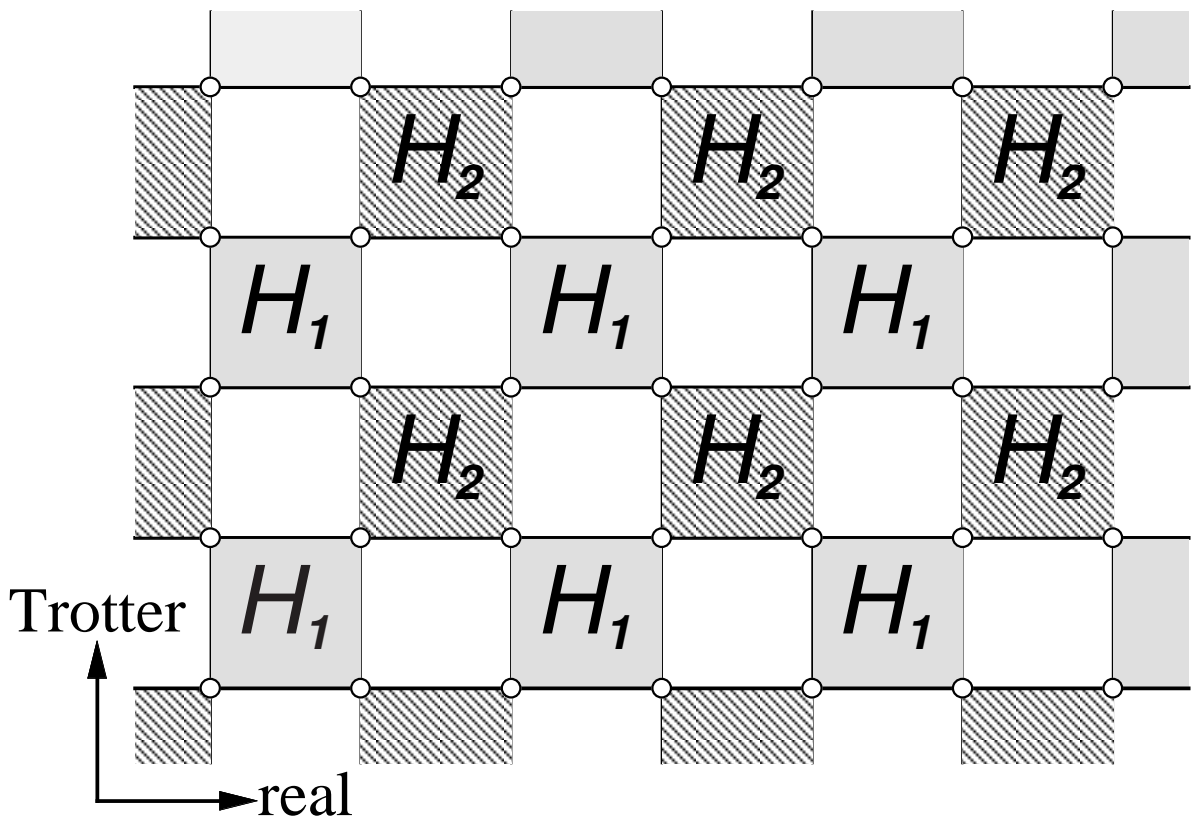
Figure 1: The checkerboard decomposition in one-dimensional systems.
\bigskip
\par
\noindent
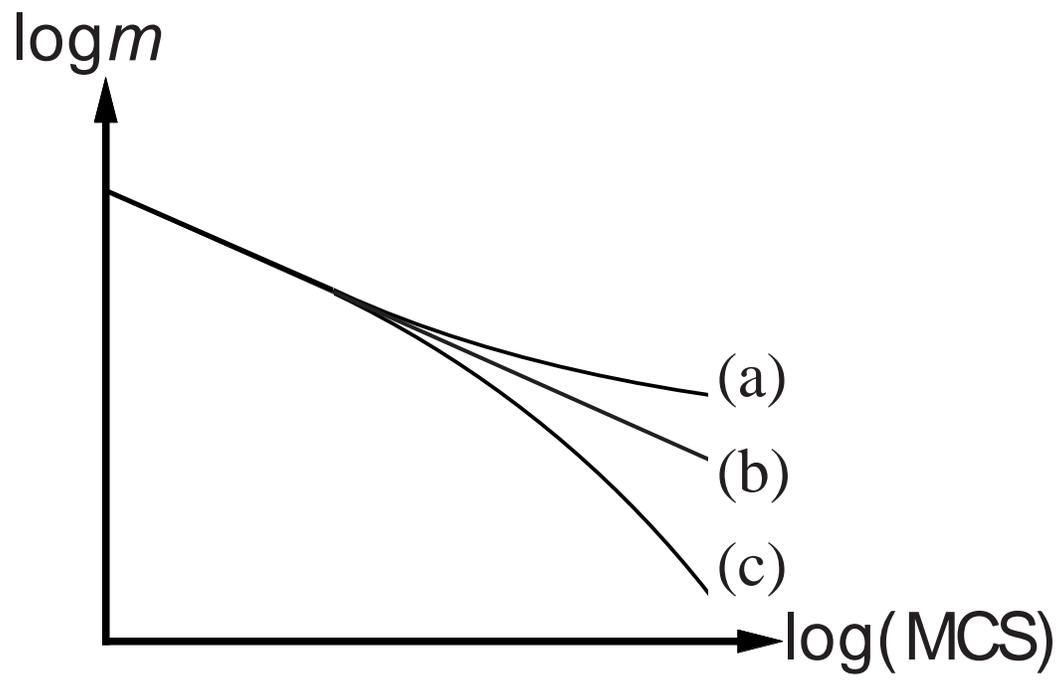
Figure 2: Schematic behavior of the non-equilibrium relaxation of 
the order parameter from a perfectly-ordered state. Values of 
the order parameter are plotted versus Monte Carlo steps 
in a logarithmic scale (a) in the ordered phase, (b) at 
the transition point, and (c) in the disordered phase. 
\bigskip
\par
\noindent
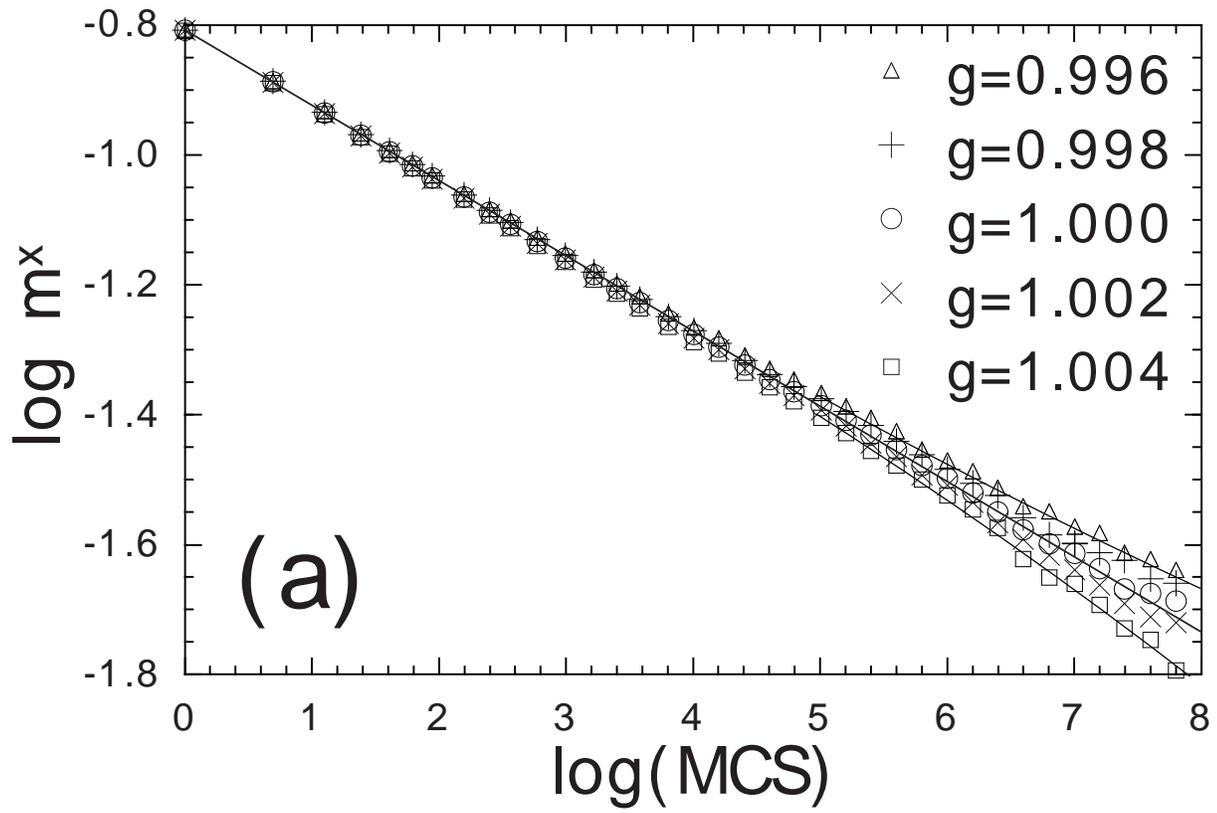
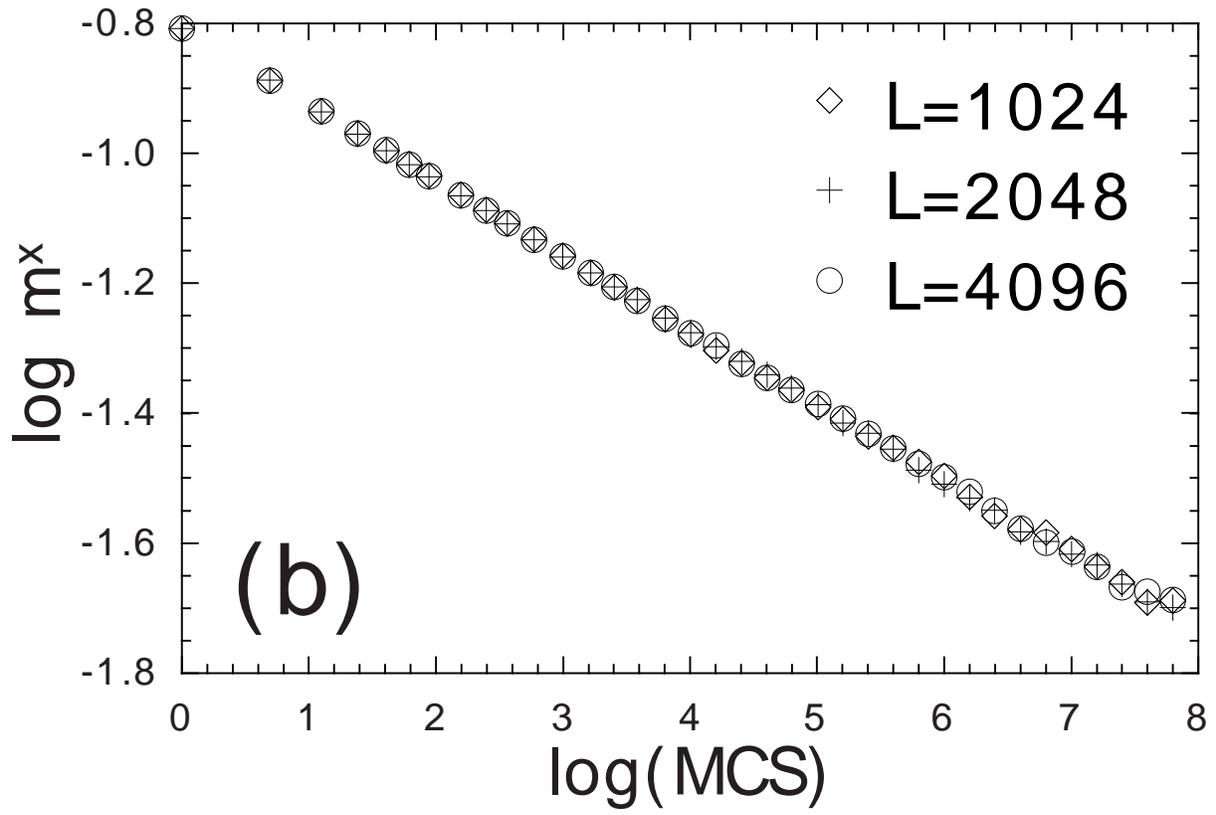
Figure 3: The order parameter of the present model is plotted 
versus Monte Carlo steps in a logarithmic scale (a) for $L=4096$ at 
$g=0.996\sim 1.004$ and (b) for $L=1024$, $2048$ and $4096$ at 
$g=1.000$. Solid lines are drawn as visual guides in the first figure. 
\bigskip
\par
\noindent
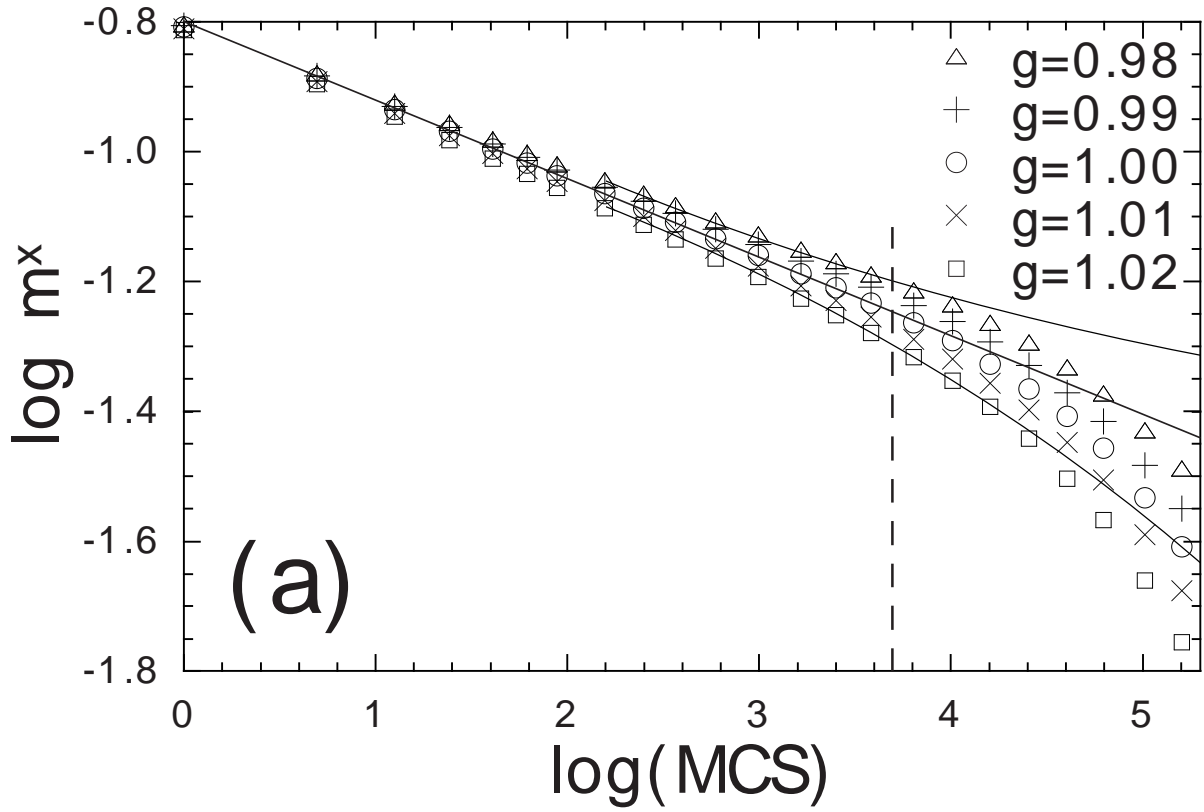
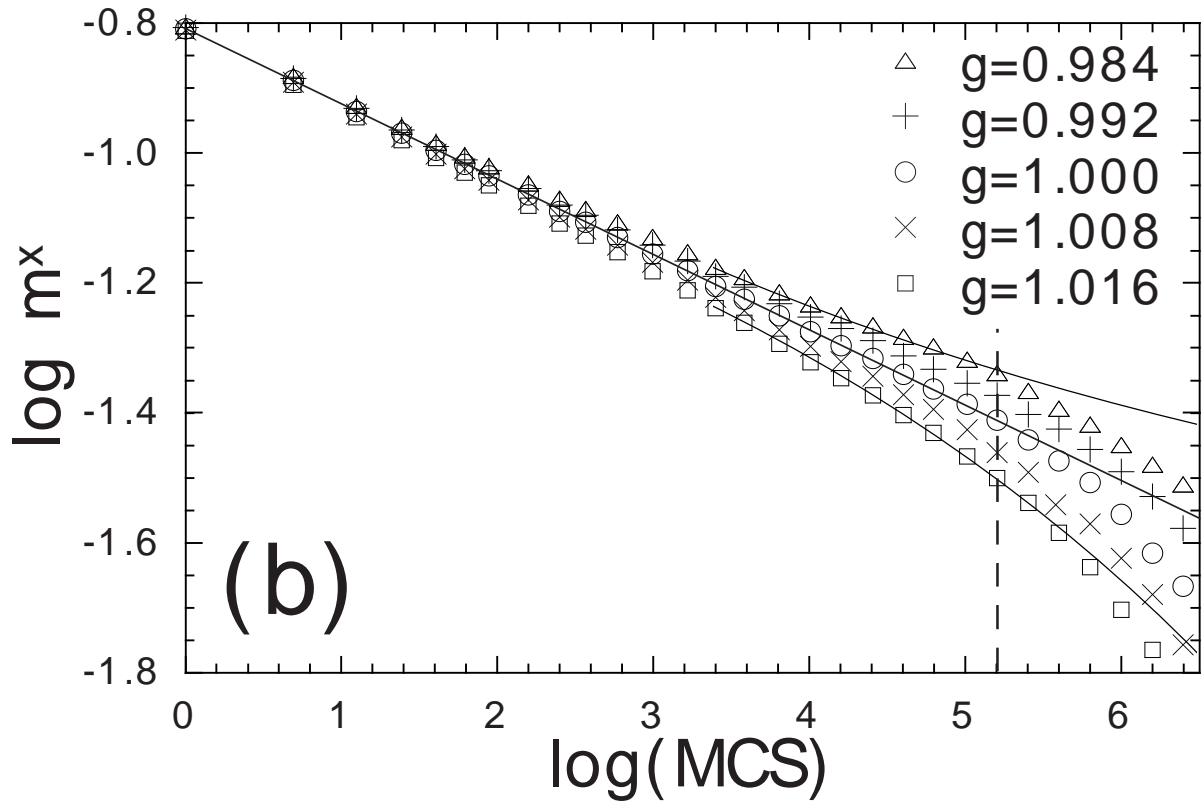
Figure 4: The order parameter of the present model is plotted versus 
Monte Carlo steps in a logarithmic scale for (a) $L=32$ and 
(b) $L=64$. The crossover from ``quasi-infinite" properties 
to finite-size properties is clearly seen. The crossover 
time scale is shown by the dashed line in each figure. 
\bigskip
\par
\noindent
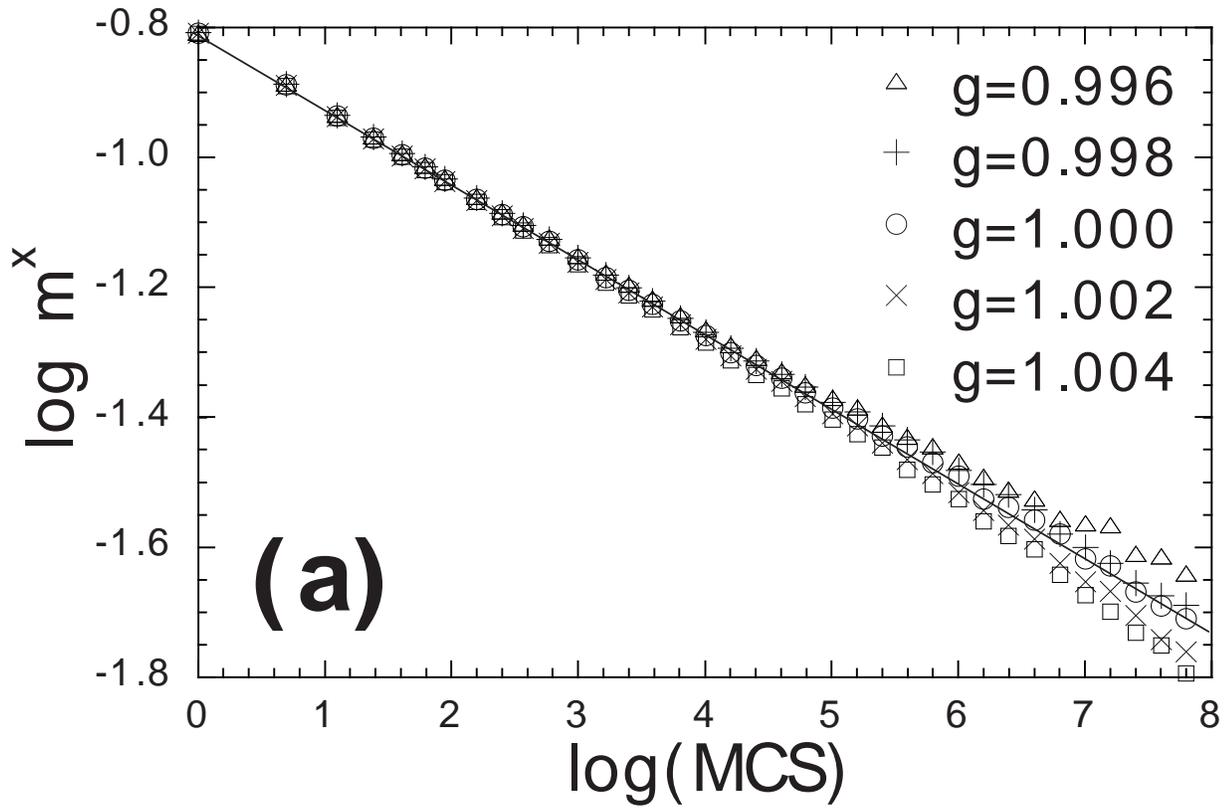
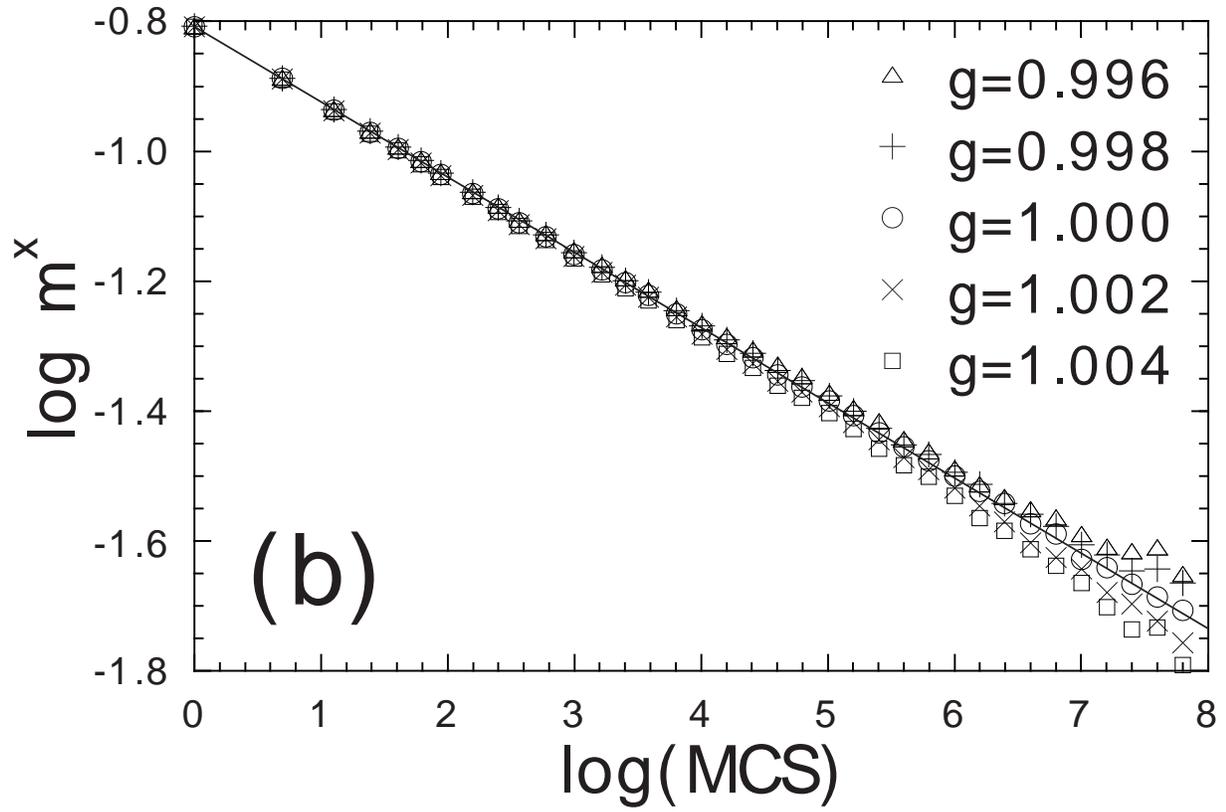
Figure 5: The order parameter of the present model is plotted 
versus Monte Carlo steps in a logarithmic scale (a) for $L=4096$ and 
$M=1024$ at $g=0.996\sim 1.004$ and (b) for $L=2048$ and $M=2048$ 
at $g=0.996\sim 1.004$. Solid lines are drawn as visual guides. 
\bigskip
\par
\noindent
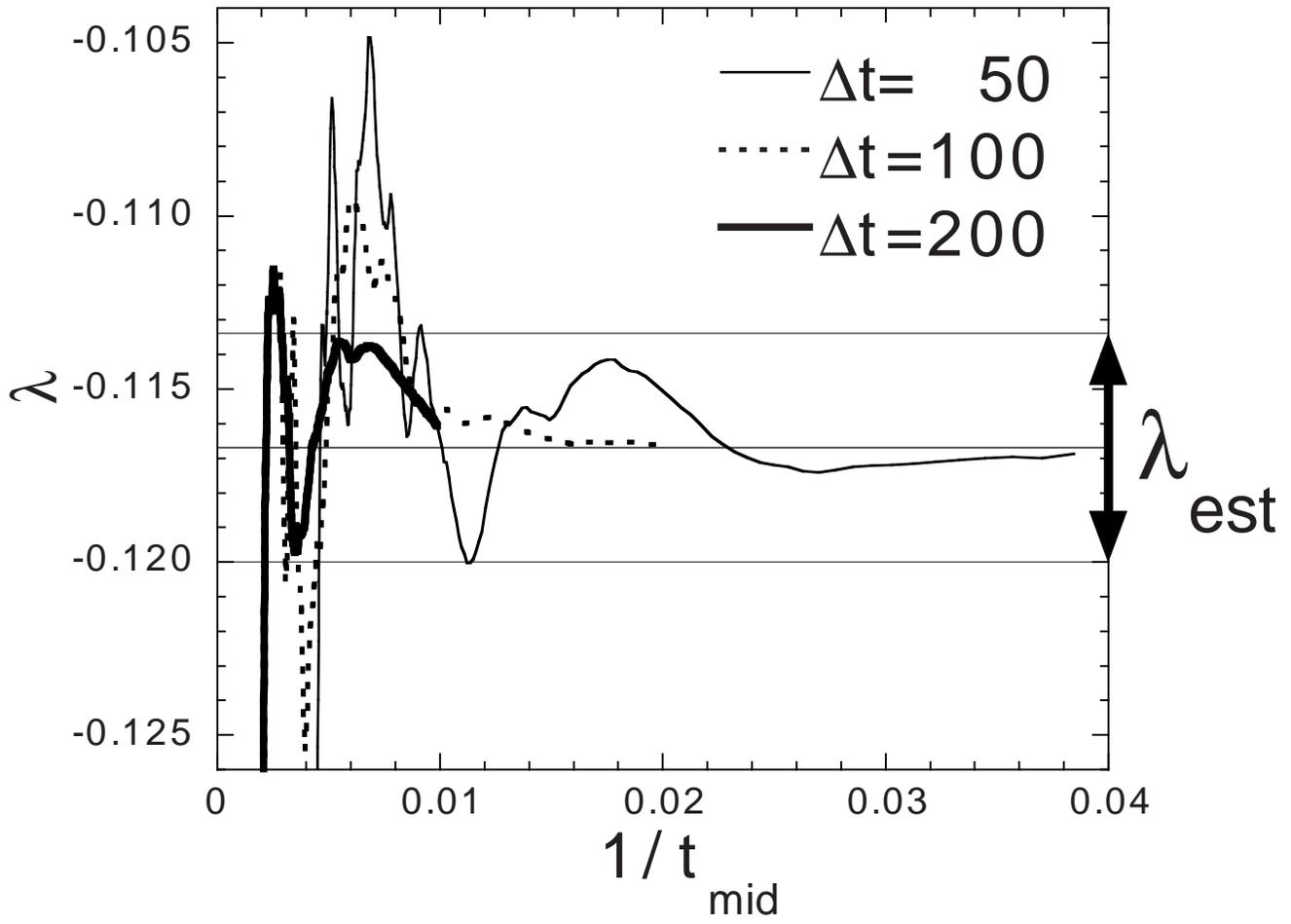
Figure 6: Estimates of the exponent $\lambda$ for various 
ranges of time scales ($t=t_{{\rm ini}}\sim t_{{\rm fin}}$, 
$\Delta t=t_{{\rm fin}}-t_{{\rm ini}}=50$, $100$ and $200$) 
are plotted versus $1/t_{{\rm mid}}$, with 
$t_{{\rm mid}}=(t_{{\rm ini}}+t_{{\rm fin}})/2$. 
The solid, dotted and bold-solid curves correspond to the 
results for $\Delta t=50$, $100$ and $200$, respectively. 
The final estimate $\lambda_{{\rm est}}=0.1167 \pm 0.0033$ 
is displayed by straight lines. 
\bigskip
\par
\noindent
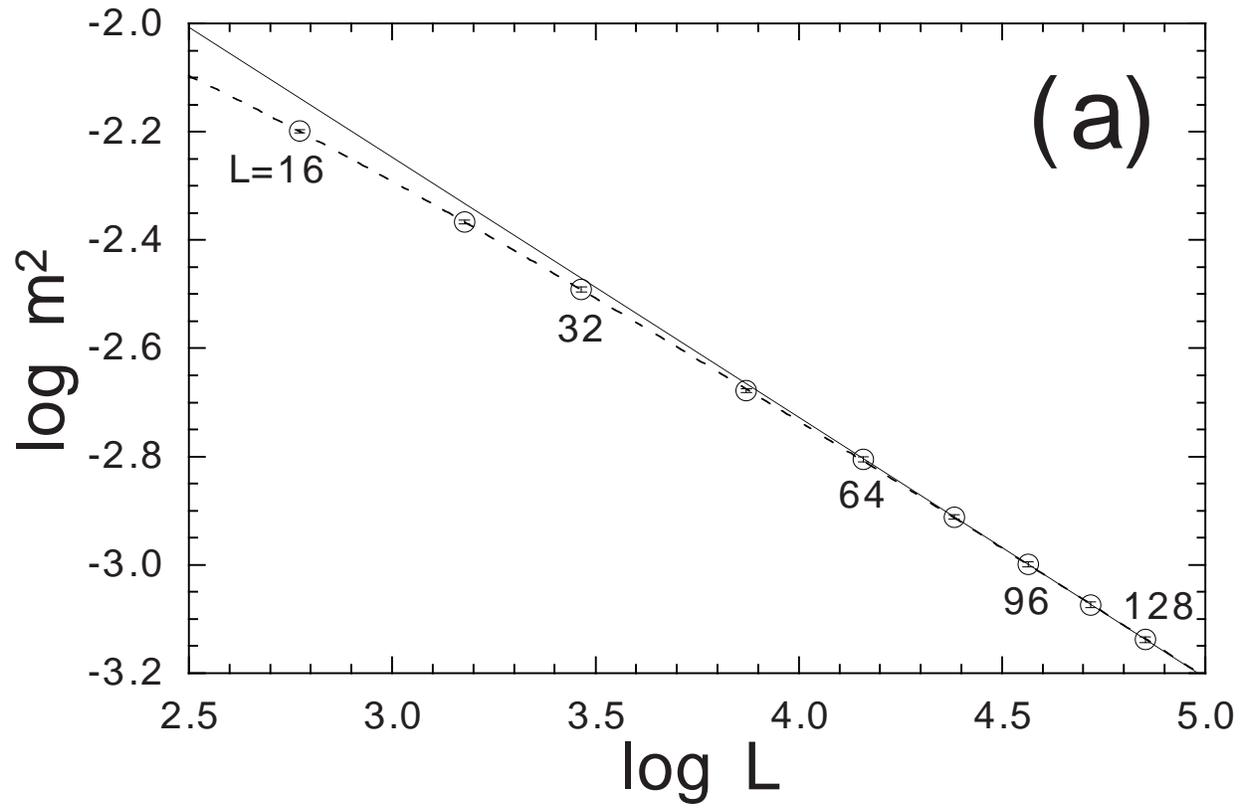
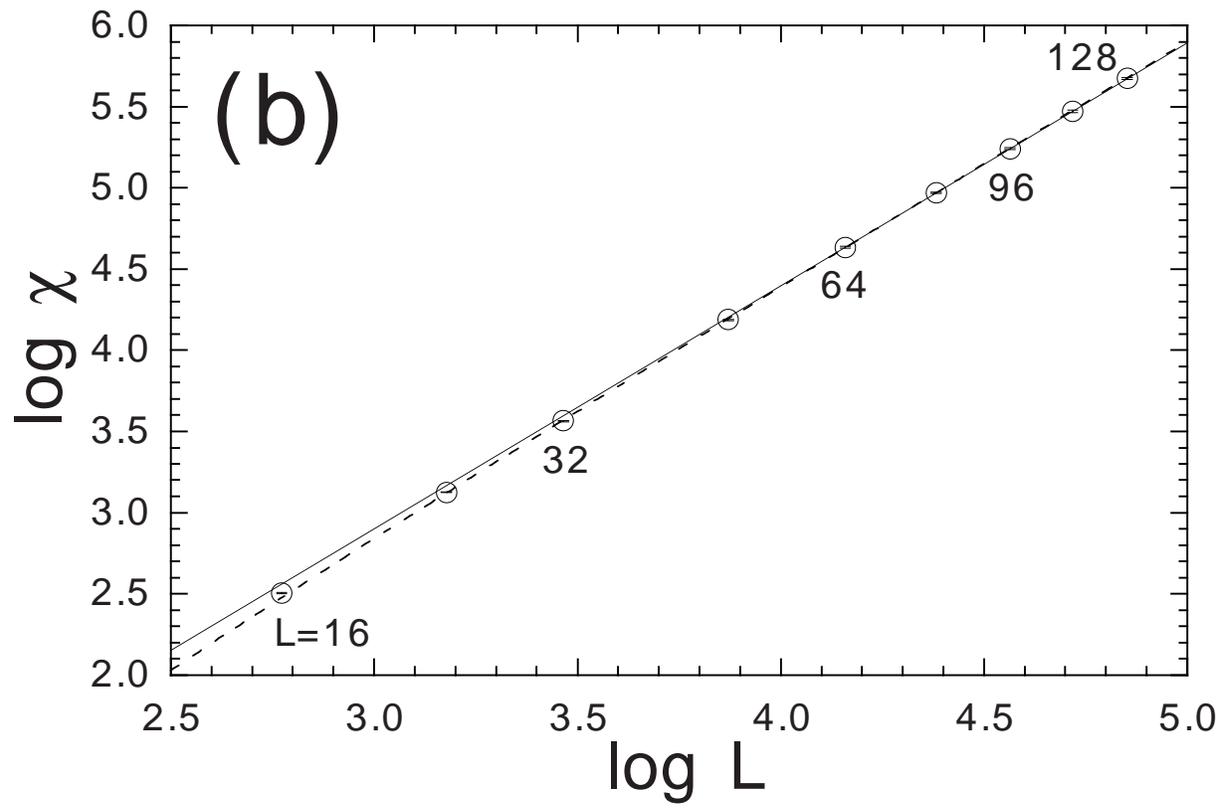
Figure 7: Ground-state averages of (a) the squared magnetization and 
(b) the susceptibility at the transition point are plotted versus system 
sizes in a logarithmic scale. The solid lines are drawn by the 
least-squares fitting of the data for $64\leq L\leq 128$ with (11), 
and the dashed lines are obtained by the least-squares fitting 
of the data for $24\leq L\leq 128$ with (16) and (17). 
\bigskip
\par
\noindent
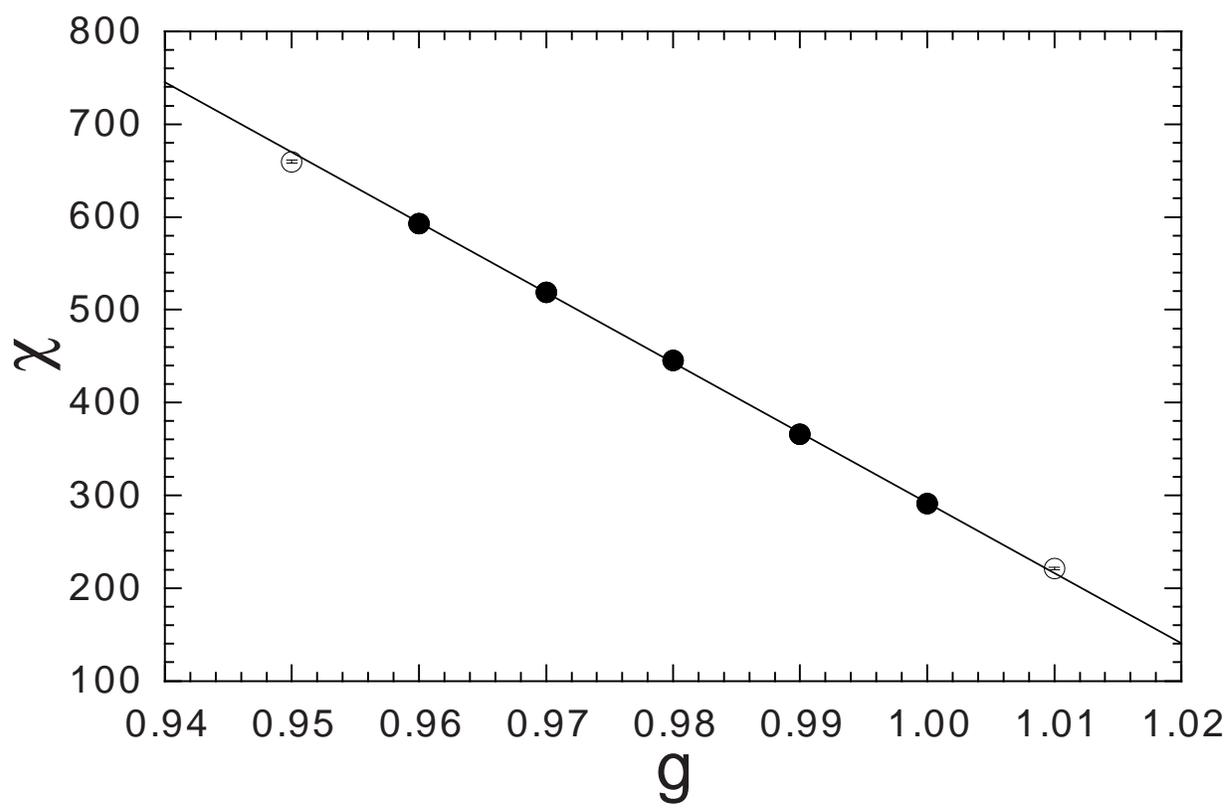
Figure 8: The susceptibility in the $L=128$ cluster is plotted versus 
the control parameter $g$ in the vicinity of the transition point. 
The slope of these data (the solid line) is evaluated by the 
least-squares fitting of the data drawn with full symbols. 
\bigskip
\par
\noindent
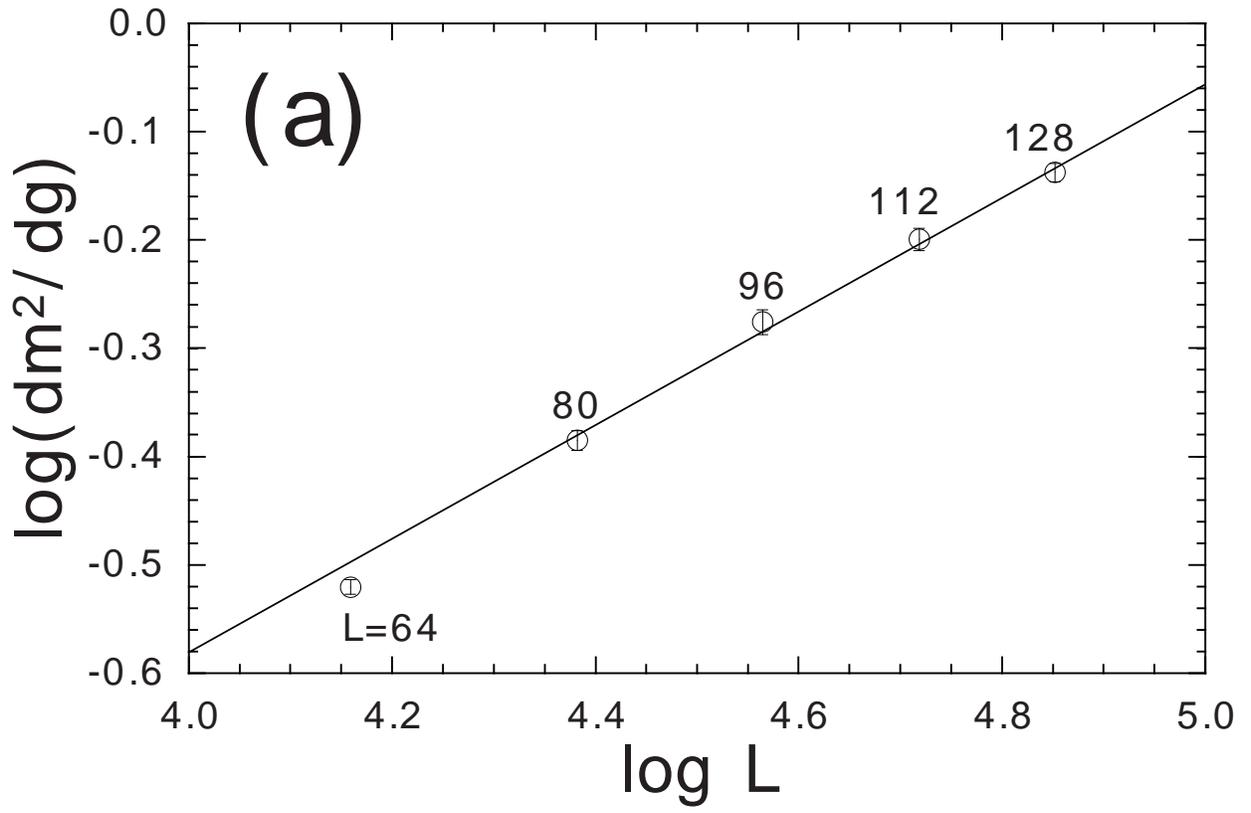
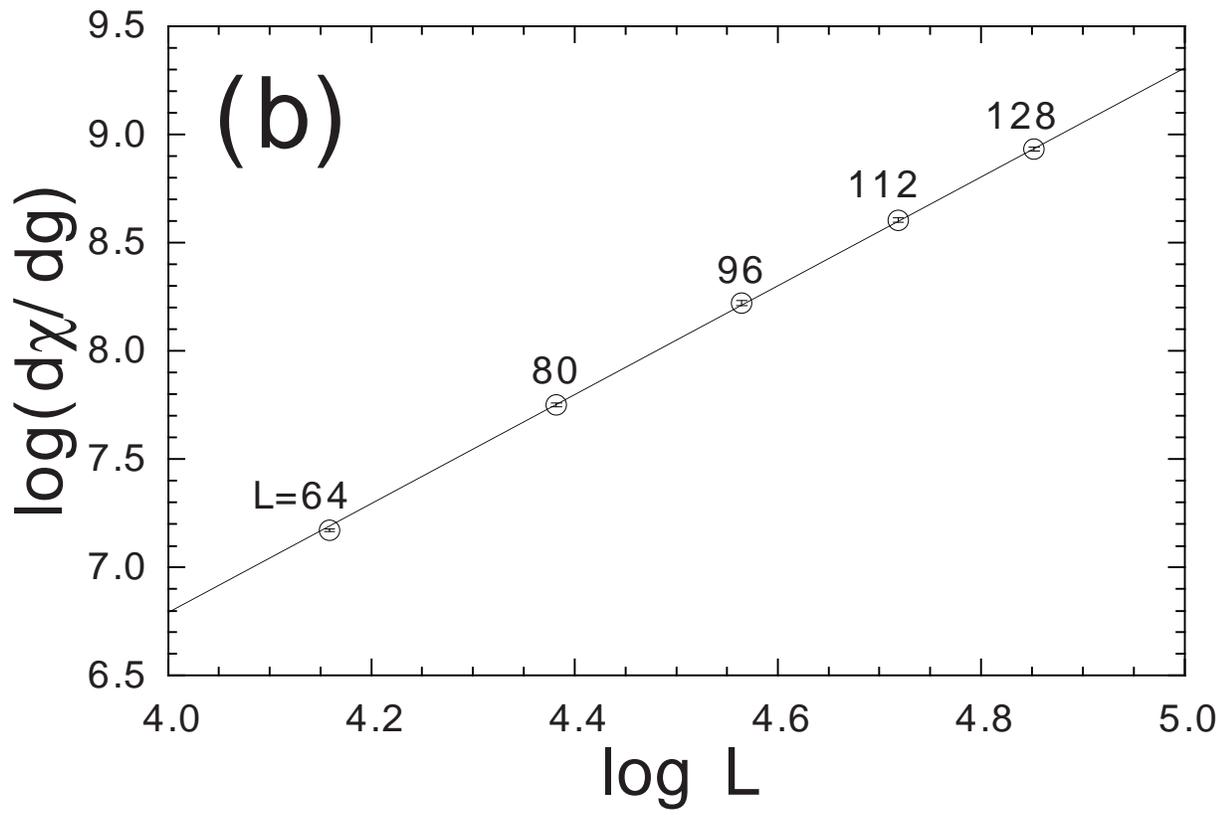
Figure 9: The slopes of (a) the squared magnetization 
and (b) the susceptibility are plotted versus system sizes in a 
logarithmic scale. The solid lines are drawn by the least-squares 
fitting of the data for $80\leq L\leq 128$ with (22). 
\bigskip
\par
\noindent
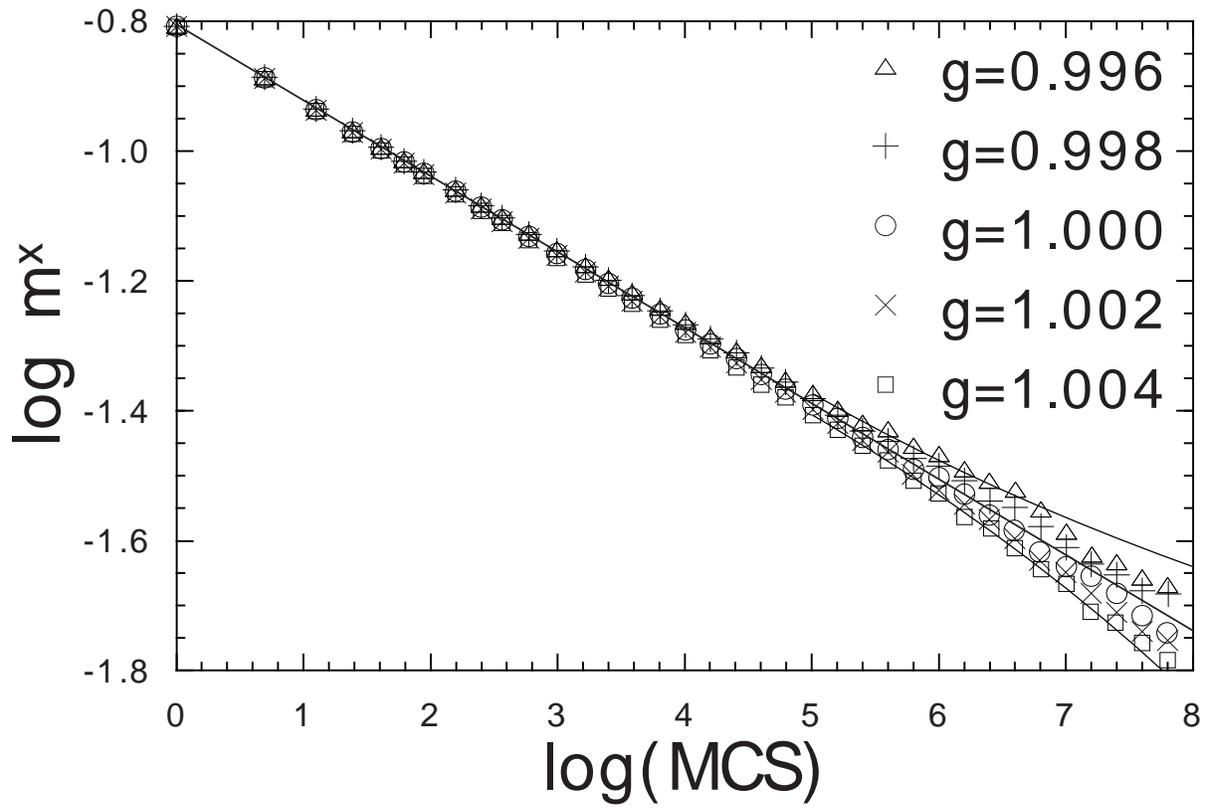
Figure 10: The order parameter of the present model including the 
Trotter-direction global flip is plotted versus Monte Carlo steps in a 
logarithmic scale for $L=4096$. Solid lines are drawn as visual guides. 
\bigskip
\par
\noindent
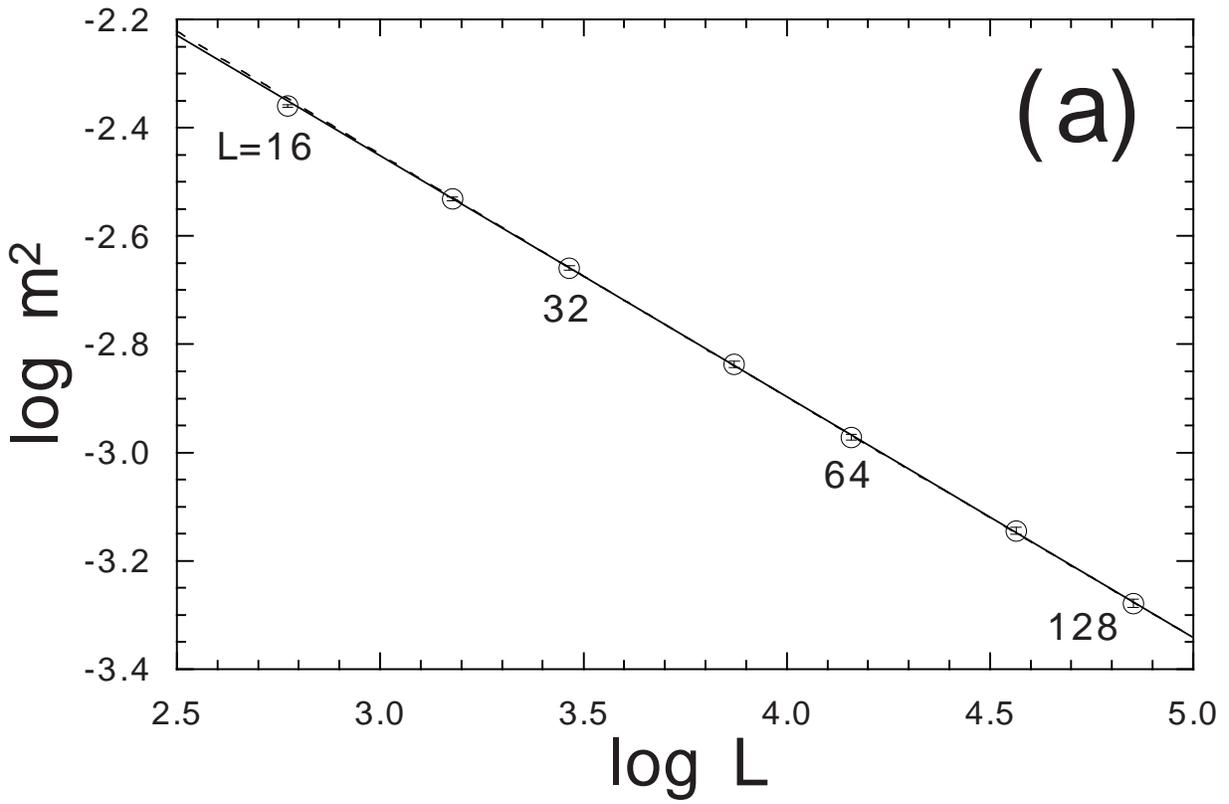
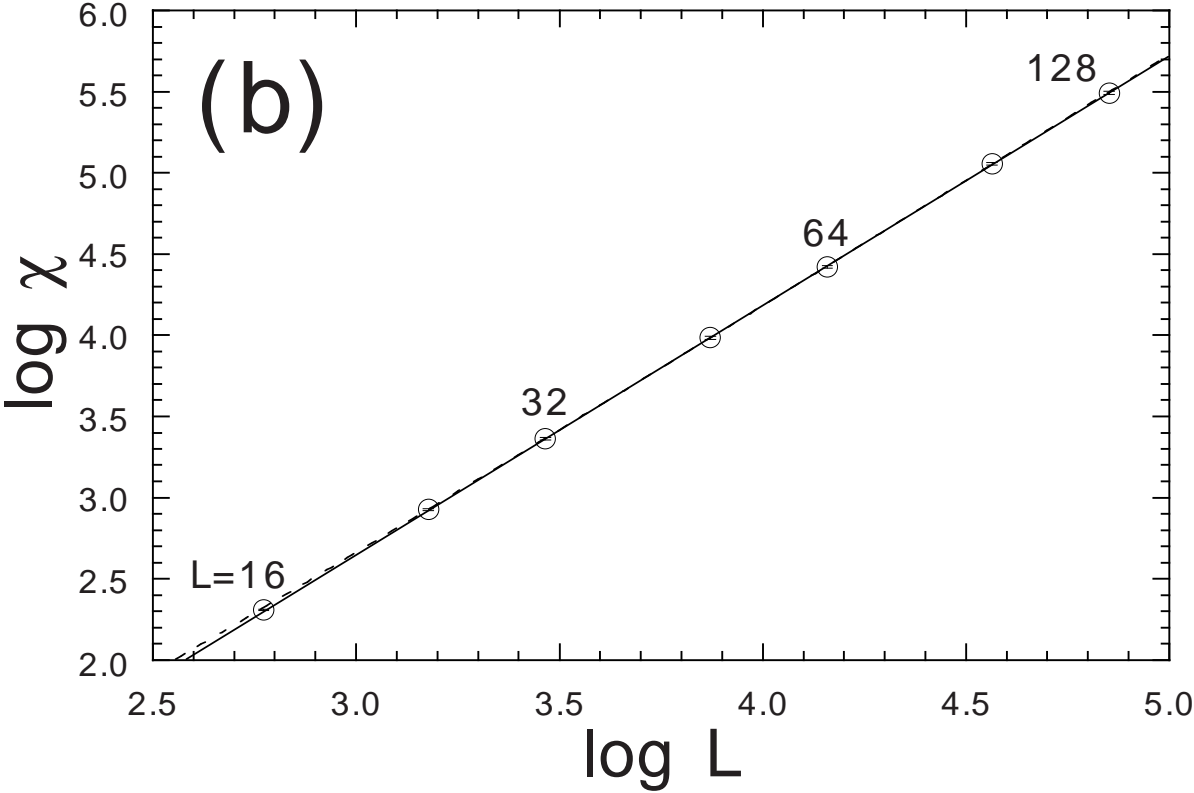
Figure 11: Size dependence of (a) the squared magnetization and 
(b) the susceptibility at the transition point in a logarithmic scale. 
The Trotter-direction global flip is included in these simulations. 
The solid lines are drawn by the least-squares fitting of the data 
for $64\leq L\leq 128$ with (11), and the dashed lines 
are obtained by the least-squares fitting of the data 
for $32\leq L\leq 128$ with (16) and (17). 

\begin{figure}
\label{decfig}
\end{figure}
\begin{figure}
\label{decschfig}
\end{figure}
\begin{figure}
\label{decsimfig}
\end{figure}
\begin{figure}
\label{decsmlfig}
\end{figure}
\begin{figure}
\label{diffig}
\end{figure}
\begin{figure}
\label{dynexpfig}
\end{figure}
\begin{figure}
\label{datafig}
\end{figure}
\begin{figure}
\label{slopefig}
\end{figure}
\begin{figure}
\label{slopefitfig}
\end{figure}
\begin{figure}
\label{decgffig}
\end{figure}
\begin{figure}
\label{scgffig}
\end{figure}
\end{document}